\documentclass[12pt]{iopart}

\usepackage{graphicx}
\usepackage{dcolumn}
\usepackage{bm}
\usepackage[utf8]{inputenc}
\usepackage{amsthm, amssymb}
\usepackage{iopams}
\usepackage{epstopdf}
\newtheorem{thm}{Theorem}

\begin{document}


\title{Nonlinear stability of the Taub-NUT soliton in 6+1 dimensions}

\author{Marek Lipert}

\address{
M. Smoluchowski Institute of Physics, Jagiellonian University, Kraków, Poland 
}

\date{\today}

\begin{abstract}
Using mixed numerical and analytical methods we give evidence that the 6+1 dimensional Taub-NUT soliton is asymptotically nonlinearly stable against small perturbations preserving biaxial Bianchi IX symmetry. We also show that for sufficiently strong perturbations the soliton collapses to a warped black hole. Since this black hole solution is not known in closed form, for completeness of the exposition we prove its existence and determine its properties. In particular, the mass of the black hole is computed. 
\end{abstract}

\maketitle

\section{Introduction}
It has been shown by Bizoń, Chmaj and Schmidt \cite{biz1}  that in odd number of spacetime dimensions one can perform a consistent 
cohomogeneity-two symmetry reduction of the vacuum Einstein equations which – in contrast to the spherically 
symmetric reduction – admits time dependent solutions. The key idea was to modify the standard 
spherically symmetric ansatz by replacing the round metric on the $S^n$ with the homogeneously squashed 
metric. By squashing we mean here the construction of a metric for $S^n$ that does not admit full symmetry group $SO(n+1)$. Instead, the metric is required to be invariant under transitive action of a specific subgroup of $SO(n+1)$. This idea, technically called  the BCS ansatz, has been used for the stability analysis of the 4+1 dimensional Taub-NUT soliton, also known as the Kaluza-Klein (KK) monopole \cite{biz2}. The monopole is important in M/String theory, as it is super-symmetric and can be used as a base for construction of certain BPS states \cite{struny,sol}. The KK monopole was found to be stable within the ansatz (for small perturbations). 

One might wonder why the KK monopole is stable. In order to address this issue we decided to change dimensionality of the model. This way we could figure out which features are dimension dependent. We chose to go from the 4+1 to the 6+1 spacetime dimensions. There is an additional advantage of this approach connected with the fact that the 6+1 dimensional Taub-NUT soliton (TN) is not super-symmetric. Therefore we can also see which aspects of relaxation rely on super-symmetry.
We found that introduced changes do not destroy stability, as the TN soliton is nonlinearly stable against small finite-energy perturbations within the BCS ansatz.
Our techniques used for stability investigation follow (and extend) the ones used in \cite{biz2}, therefore the reader is encouraged to read both papers in conjunction.

Further research revealed that strongly perturbed TN soliton loses stability and collapses, forming a black hole. This also occurred in the lower dimensional model. This time, however, the black hole solution is not known in closed form. We examined it in detail, proving its existence and computing its asymptotic expansion. This allowed us to compute its mass and compare it to the mass of its lower dimensional cousin.

\section{Taub-NUT soliton}
The Taub-NUT soliton in 6+1 dimensions is a static regular nontrivial solution
of the vacuum Einstein equations. 
It can be explicitly found in \cite{sol}:
\begin{eqnarray}
\label{soliton}
ds^2_{6+1} &=& -dt^2 + \frac{(\rho+2 l)^2}{2\rho(\rho+4l)}d\rho^2 + \rho(\rho+2l) d\Sigma^2_2+ \frac{2l^2\rho(\rho+4l)}{(\rho+2 l)^2} \sigma^2 ,
\end{eqnarray}
where $l$ is a free positive parameter, 
\begin{eqnarray}
\label{fubini}
\sigma^2 &=& (d\mu - \frac{1}{2}\sin^2{\xi}\sigma_3)^2, \\
\nonumber d\Sigma^2_2 &=& d\xi^2 + \frac{1}{4} \sin^2{\xi}~(\sigma^2_1 + \sigma^2_2 + \cos^2{\xi} ~\sigma^2_3),
\end{eqnarray}
the latter being the Fubini-Study metric for $\mathbb{CP}^2$ and
\begin{eqnarray*}
\label{sigmy}
\sigma_1 &=& \sin{\psi}\sin{\theta} d \phi  + \cos{\psi} d\theta,\\
\sigma_2 &=& \cos{\psi}\sin{\theta} d\phi - \sin{\psi} d\theta,\\
\sigma_3 &=& d \psi + \cos{\theta} d\phi
\end{eqnarray*}
are standard Maurer-Cartan forms for $SU(2)$. Parameter ranges are as follows 
\begin{eqnarray*}
0 \leq \rho, \quad 0 \leq \phi \leq 2 \pi,  \quad 0  \leq  \theta \leq  \pi,\\ 
0 \leq \psi \leq 4\pi, \quad    0 \leq \xi \leq \frac{\pi}{2},\quad  0 \leq \mu \leq 2\pi. 
\end{eqnarray*}

It is worth noticing that this solution lacks asymptotic flatness in a fashion similar to its 4+1 dimensional counterpart. It 
approaches $\mathbb{R}^5$ times a circle of asymptotic radius $\sqrt{2}l$ near infinity.  Near $\rho=0$, on the other hand, it is locally flat, which can be seen by introducing a new radial coordinate $s = \sqrt{2l \rho}$.
Then the solution is interpolating between two different vacua which justifies the name: soliton.

\section{Preliminaries}
The squashed $S^5$ can be constructed as a coset $SU(3)/SU(2)$ with $SU(3) \subset SO(6)$ being
its symmetry group. The result is widely known and can be easily derived using the methods of \cite{besse,wraith,cartan,parametr}. The BCS ansatz has the form
\begin{equation}
\label{ansatz}
ds^2 = -Ae^{-2\delta}dt^2 + A^{-1}dr^2 + r^2 (e^{-8 B} \sigma^2 + e^{2 B} d\Sigma^2_2),
\end{equation}
where $\sigma^2$ and $d\Sigma^2_2$ were defined in \eref{fubini} and $A$, $B$, $\delta$ are functions of
$r$ and $t$. The function $B$, which measures squashing resulting from lower symmetry of 
our description of $S^5$, serves as a dynamical degree of freedom and breaks
Birkhoff's theorem for that case. It has been chosen in such a way that it does not change spheres' volume. 

Substituting the metric (\ref{ansatz}) into the vacuum Einstein equations we get the following system of equations
\numparts
\begin{eqnarray}
\label{prozniowe}
&A'=-\frac{4 A}{r}+\frac{1}{5r}(24e^{-2B}-4e^{-12B})-4r(e^{2\delta}A^{-1}\dot{B}^2+AB'^2),\\
&\dot{A} = -8rA\dot{B}B',\\
&\delta' = -4r(e^{2\delta}A^{-2}\dot{B}^2 +B'^2),\\
\label{ostatnie}
&\left(e^\delta A^{-1}r^5 \dot{B}\right)^{^.}-\left(e^{-\delta}Ar^5B'\right) '+\frac{6}{5}e^{-\delta}r^3(e^{-2B}-e^{-12B})=0,
\end{eqnarray}
\endnumparts
where primes and dots denote derivatives with respect to $r$ and $t$, respectively. Note that the equations are scale invariant.
The Taub-NUT solution (\ref{soliton}) expressed in terms of ansatz (\ref{ansatz})
takes the form
\begin{eqnarray}
\label{solitonx}
e^{-10B_0} = \frac{2l^2(\rho+4l)}{(\rho+2l)^3}, &\quad A_0 = \frac{2(\rho+4l)\rho}{(\rho+2l)^2} (\frac{dr}{d\rho})^2,&\quad e^{2\delta_0} = A_0, 
\end{eqnarray}
\begin{eqnarray}
\label{riro}
&r^{10} = 2l^2\rho^5 (\rho+2l)^2 (\rho+4l). 
\end{eqnarray}
It can be proven that every static regular solution of (\ref{prozniowe}-\ref{ostatnie}) should exhibit 
Taylor series expansion near $r=0$
\begin{equation}
\label{wokolzera}
 B(r) = br^2+O(r^4),\; \;  A(r) = 1-8b^2 r^4 + O(r^8), \; \; \delta(r) = -4b^2r^4 + O(r^6)
\end{equation}
with $\sqrt{b}$ being the scale factor (a rigorous proof can be found in appendix A).  
It is easy to verify that (\ref{solitonx}) possesses the same expansion, with
$b=\frac{1}{16l^2}$. Thus we identify $l$ with the scaling freedom and conclude
that up to scaling there exists the unique solution (\ref{solitonx}).  

\section{Linear stability}
The linear stability analysis proceeds along the same lines as in the 4+1 dimensional case.
Assuming 
\begin{eqnarray}
A(r,t) = A_0(r) + A_1(r,t),\quad  B(r,t) = B_0(r) +  B_1(r,t), \quad\\
\nonumber \delta(r,t)=\delta_0(r)+\delta_1(r,t),
\end{eqnarray}
and keeping only linear terms in perturbations, we get an equation for a single Fourier
mode ($\rho(r)$ is defined by equation \eref{riro}) , $B_1(\rho(r),t)=v(\rho) e^{\rmi \lambda t}$: 
\begin{equation}
\label{strum}
L v(\rho) = \lambda^2 v(\rho),
\end{equation}
with
\begin{eqnarray}
L =  \frac{-2}{\rho^2(\rho+2l)^2} \frac{\rmd }{\rmd \rho} \left(\rho^3(\rho+4l) \frac{\rmd}{\rmd \rho}  \right) + V(\rho).\nonumber
\end{eqnarray}
where $V(\rho)$ is a long and complicated expression (given in the appendix B). 
This equation cannot be solved analytically for generic $\lambda^2$.
However, in order to demonstrate linear stability we only need to know the zero mode. This mode can be deduced from the scaling freedom. For $b$ near 1 we have
\begin{equation}
B_0(b r) = B_0(r) + r \frac{\rmd}{\rmd r} B_0(r) (b-1) + O((b-1)^2). \nonumber 
\end{equation}
We define $ v_0 = r \frac{\rmd}{\rmd r} B_0(r)$. Changing $r \to \rho$ we get
\begin{equation}
\label{zeros}
v_0(\rho) = {\frac { \left( \rho+5\,l \right) \rho}{4({\rho}^{2}+5\,\rho\,l+5\,{l}^{2})}},
\end{equation}
with $0\leq\rho$, $0<l$. It can be easily verified that this solves (\ref{strum}) for $\lambda^2=0$. We see that the zero mode is manifestly positive for all $\rho$ in 
the domain. This implies by the standard  Sturm-Liouville theory that the operator $L$ has no negative eigenvalues. Thus, the TN soliton is linearly stable within our ansatz. Note, that the zero mode is not a genuine eigenfunction because it is not square integrable. Rather, it is a resonance sitting at the bottom of the continuous spectrum.

\section{Warped black holes}
Black hole solutions play the key role in dynamics of a strongly perturbed Taub-NUT soliton. 
As far as we know, in contrast to the 4+1 dimensional case, these solutions are not known in closed form in 6+1 dimensions. Thus, from the mathematical point of view the very existence of such solutions is not obvious and requires a proof. Following Breitenhohner, Forg\'{a}cs and Maison \cite{proof} one can prove the local existence (see Appendix A). Global existence proof, as well as analysis of an asymptotic behaviour for large $r$ is described below. 

Let us begin by explaining what we mean by a black hole solution in our model. If we assume that for a given $r_H>0$ $A(r_H)=0$ and substitute the formal power series expansion into the static equations, we obtain:
\begin{eqnarray}
\label{rozw2}
&B(r) \sim \beta + \frac{6}{4}\frac{e^{-2\beta}-e^{-12\beta}}{6e^{-2\beta}-e^{-12\beta}}\left(\frac{r}{r_H} - 1 \right),\\
&\nonumber A(r) \sim \frac{4}{5} (6 e^{-2 \beta}-e^{-12 \beta})\left(\frac{r}{r_H}-1\right).
\end{eqnarray}
Theorems presented in the Appendix A guarantee that a local solution with the above behaviour for small $r-r_H$ exists and is analytic in $\beta, r_H$ and $r-r_H$. By setting $\beta=0$ one arrives at the ordinary Schwarzschild black hole with $B=0$. Numerical simulations show that solution with $\beta<0$  plays no role in TN soliton evolution. 
Below, the global solution with local behaviour \eref{rozw2} and $\beta>0$ will be called the warped black hole (WBH).

\subsection{Global properties}
We now focus on establishing the global existence of WBH.
Using logarithmic variable $\tau = \ln{r}$, static version of the equations (\ref{prozniowe}-\ref{ostatnie}) can be cast into the following form:
\numparts
\begin{eqnarray}
\label{auto1}
A B''  + \frac{1}{5} \left[ B' \left(24 e^{-2B} - 4 e^{-12B}\right) - 6 \left(e^{-2B} - e^{-12B} \right) \right] = 0, \\
\label{auto2}
A' = -4A\left(1+B'^2\right) + \frac{1}{5}\left( 24e^{-2B}-4 e^{-12B} \right), \\
\label{auto3}
\delta' = -4B'^2.
\end{eqnarray}
\endnumparts
Here prime corresponds to differentiation with respect to $\tau$.
This is an autonomous system, which is a consequence of scaling invariance.
Note that the first two equations are independent of the third one.

It can be easily verified that WBH solution obeys the following inequalities in a right-open range $[\tau_H,\tau_1)$  where $\tau_H=\ln{r_H}$:
\begin{equation}
\label{dobre}
0<B'<\frac{1}{4},\quad 0<A<1,\quad 0<B.
\end{equation}
Our argument proceeds in two steps. In the first step we show that if the local WBH solution  preserving \eref{dobre} is given in some right-open interval $I = [\tau_0,\tau_+)$, then the solution can be safely extended to $\bar{I}=[\tau_0,\tau_+]$ preserving \eref{dobre}. In the second step we prove that the first step is equivalent to global existence of the WBH.

\subsection*{Step I}
Let us define an auxiliary function
\begin{eqnarray}
f(B)=\frac{6(e^{-2B}-e^{-12B})}{24e^{-2B}-4e^{-12B}}.
\end{eqnarray}
and observe, that $\frac{1}{4}-\epsilon>f(B)$ for $r \in I$ and some finite $\epsilon>0$ dependent on the choice of $I$.
Now, if we take $B''<0$ and look at (\ref{auto1}) we get
\begin{eqnarray}
\label{podstawa}
B' > f(B), 
\end{eqnarray}
for $r \in I$. Therefore, we can safely assume that $B''<0$ for $B'>\frac{1}{4}-\epsilon$.
This naturally leads to a conclusion that $B'(\tau_+)<\frac{1}{4}$.
Next, we can put $B'=0$ in (\ref{auto1}) to obtain
\begin{equation}
\nonumber B''=\frac{6}{5A}(e^{-2B}-e^{-12B})>0 ,
\end{equation}
which means that $B'(\tau_+)>0$.  We can proceed with similar considerations for the function $A$. Setting $A=1$ in (\ref{auto2}) we get  
\begin{equation}
\nonumber A' = -4 B'^2 - \frac{1}{5}\left( 20 + 4 e^{-12B} -24e^{-2B}\right) < 0 ,
\end{equation}
which means that $A(\tau_+)<1$. Finally, setting $A=0$ in (\ref{auto2}) we get
\begin{equation}
\label{asymptota}
A'=\frac{1}{5}(24 e^{-2B}-4 e^{-12B}) > 0,
\end{equation}
which means that $A(r_+)>0$. One should not be concerned with setting $A=0$ in $-4A(1+B'^2)$ because $B'$ is finite in $\bar{I}$. 
Finally, from $B'>0$ in $\bar{I}$ we get that $B>0$ in $\bar{I}$. Note, that $B(\tau_+)< \infty$  because $B'(\tau_+)<\infty$.
\subsection*{Step II}
Let us assume that a local WBH solution satisfying \eref{dobre} in some $[\tau_H,\tau_+)$ cannot be extended to infinity preserving \eref{dobre}.
Then, there exists $\tau_s\geq \tau_+$ where  \eref{dobre} is not satisfied. By our construction, the solution must be finite and preserve \eref{dobre} for $I=[\tau_+,\tau_s)$. We can now use the results of \textit{Step I} to get a contradiction - the solution is regular and preserves \eref{dobre} in $\tau_s$.  
\\

We shall now prove that $B \sim \tau$ for large $\tau$.
To see this, let us first note that $B(\tau) \to \infty$ as $\tau \to \infty$. Indeed, we already know from \eref{dobre} that $B$ is monotonically increasing. Furthermore, if it was to reach a finite limit as $\tau$ grows, $B'$ would have to approach $0$ and $f(B(\tau))$ would attain a finite positive limit $\lim_{\tau \to \infty}{f(B(\tau))}=\xi>0$. However, if we reverse a sign in \eref{podstawa} we immediately arrive at the contradiction because whenever $B'$ drops below $\xi$, $B''$ becomes positive repelling $B'$ from $0$.
 
Next, let us observe that
\begin{eqnarray}
 \frac{ \rmd}{\rmd \tau} f(B(\tau)) = B' f'(B) > 0.
\end{eqnarray}
In the neighbourhood of $\tau_H = \ln{r_H}$ we have $B'=f(B)$ which follows from \eref{rozw2}. As we depart towards greater $\tau$, $B'$ must drop below $f(B)$ because from \eref{podstawa} we know that initially $B''=0 < \frac{ \rmd}{\rmd \tau} f(B(\tau))$. It becomes evident, by the same argument, that once $B'$ drops below $f$ it can never cross it again. Therefore $B'$ is monotonically increasing (from reversed \eref{podstawa}) and bounded from above (because $f$ is bounded from above and $B'<f$). We get $\lim_{\tau \to \infty} B'(\tau) = \zeta>0$, which implies
\begin{equation}
\label{asymptot}
B(\tau) \sim \zeta \tau + \Sigma
\end{equation}
for large $\tau$. $\Sigma$ is an integration constant that, as we shall see, corresponds to the scale invariance.

\subsection{Exact asymptotic behaviour}
We shall now focus on establishing  the large $\tau$ series expansion for WBH solutions. 
First, notice that substituting \eref{asymptot} into \eref{auto1} and neglecting $e^{-12B}$ yields the consistency condition
\begin{eqnarray}
(4\zeta - 1)e^{-2B} = 0
\end{eqnarray}
which fixes $\zeta$ at $\frac{1}{4}$. Let us now rewrite $B(\tau)=\frac{1}{4}\tau + \tilde{B}(\tau)$.
From \eref{asymptota} and \eref{auto2} we expect that $A \to 0$ as $\tau \to \infty$. We also have $\tilde{B} \to 0 $ as $\tau \to \infty$. We shall now solve the system (\ref{auto1}-\ref{auto2}) through successive approximations. From the above assumptions we get for \eref{auto2}:
\begin{eqnarray}
\label{lepsze}
 A' + \frac{17}{4}A=   \frac{24}{5} e^{-\frac{1}{2}\tau} (1-2\tilde{B}+\frac{(2\tilde{B})^2}{2!} -\ldots) - \\
\nonumber - \frac{4}{5} e^{-3\tau}(1-12 \tilde{B} + \frac{(12 \tilde{B})^2}{2!}- \ldots) - 2A\tilde{B}' - 4\tilde{B}'^2.
\end{eqnarray}
We expect that $A$ as well as $\tilde{B}$ will be expressed as a series of $e^{- \mu_n \tau}$-like expressions for subsequent $\mu_n$'s so that they eventually all cancel each other. The first exponent to be cancelled is $e^{- \frac{1}{2}\tau}$ - therefore, neglecting higher order terms, we get
\begin{equation}
A' +\frac{17}{4}A=   \frac{24}{5}e^{-\frac{1}{2} \tau}
\end{equation}
with a solution 
\begin{eqnarray}
\label{alphas}
A(\tau)=\frac{32}{25}e^{-\frac{1}{2}\tau} + \alpha e^{-\frac{17}{4} \tau},
\end{eqnarray}
where $\alpha$ comes as an integration constant for homogeneous equation. We can now rewrite \eref{lepsze} using $A(\tau)=\frac{32}{25}e^{-\frac{1}{2}\tau} + \alpha e^{-\frac{17}{4} \tau}+\tilde{A}(\tau)$ to get:
\begin{eqnarray}
\label{lepsze2}
\fl \tilde{A}'+\frac{17}{4}\tilde{A}  =  \frac{24}{5} e^{-\frac{1}{2}\tau} (-2\tilde{B}+\frac{(2\tilde{B})^2}{2!} -\ldots) 
 - \frac{4}{5} e^{-3\tau}(1-12 \tilde{B} + \frac{(12 \tilde{B})^2}{2!}- \ldots)\\ \nonumber  - 2\alpha e^{-\frac{17}{4}\tau} \tilde{B}' - \frac{64}{25}e^{-\frac{1}{2}\tau}\tilde{B'} - 2\tilde{A}\tilde{B}' - 4\tilde{B}'^2.
\end{eqnarray}
Note, that $\tilde{B}$ can not have a $e^{-\frac{1}{2}\tau}$ term because there would be no possibility to cancel $\frac{24}{5}e^{-\frac{1}{2}\tau}(-2\tilde{B})$. Instead, this factor must combine to $\e^{-3\tau}$ in order to cancel $-\frac{4}{5}e^{-3\tau}$. We can write 
\begin{equation}
e^{-\frac{1}{2}\tau}( -\frac{48}{5} \tilde{B} -\frac{64}{25}\tilde{B}') - \frac{4}{5}e^{-3\tau} = 0
\end{equation}
and solve it
\begin{equation}
\tilde{B} = -\frac{1}{4}e^{-\frac{5}{2}\tau} + \gamma e^{-\frac{15}{4}\tau}.
\end{equation}
Unfortunately, this is a wrong solution. One can see this by substituting it into \eref{auto1} and trying to find next order of expansion - unbalanced exponents appear. 
This leads to a conclusion that $A$ must also contribute to $e^{-3\tau}$ equation. It can be consistently done only by including $e^{-3\tau}$ term in $A$ as well as $e^{-\frac{5}{2}\tau}$ term in $B$:
\begin{eqnarray}
\nonumber B(\tau) = \frac{1}{4}\tau + C e^{-\frac{5}{2}\tau} + \gamma e^{-\frac{15}{4}\tau} + \tilde{B}(\tau), \\
\nonumber A(\tau) = \frac{32}{25} e^{-\frac{1}{2}\tau} + D e^{-3\tau} + \alpha e^{-\frac{17}{4}\tau} + \tilde{A}(\tau),
\end{eqnarray}
where $C$ and $D$ are yet unknown coefficients, $\tilde{B}$ has been redefined and $\tilde{A}$ has been defined for the next step of the method. After substitution into \eref{auto2} we get an equation for terms proportional to $e^{-3\tau}$
\begin{equation}
\label{wnio1}
\left( 25\,D+64\,C+16 \right) =0.
\end{equation}
Equation \eref{auto1} provides us with yet another equation for $e^{-3\tau}$ term
\begin{equation}
\label{wnio2}
1-4C = 0.
\end{equation}
We therefore get $C=\frac{1}{4},\quad D=-\frac{32}{25}$. Everything is consistent and we can proceed to higher orders by guessing what terms should we include to cancel constant factors in equations for further exponents. It is interesting that we never arrive at any constraints for $\alpha$ or $\gamma$. This constants explicitly appear in higher coefficients of the expansion, but remain completely undetermined. 

Our heuristic reasoning is justified by theorem from Appendix A which ensures that the above formal expansion defines well behaved three parameter family of local solutions near infinity. The theorem requires that we interpret the integration constant $\Sigma$ from \eref{asymptot} as a scaling constant and that we parameterize solutions with $\{\alpha,\gamma,\Sigma\}$. Below we present an expansion computed to one more order, translated into the original variable $r=e^\tau$:
\begin{eqnarray}
\label{wokolnieskonczonosci}
&A(r) \sim \frac{32}{25}\frac{1}{\sqrt{R}}-\frac{32}{25}\frac{1}{R^3}+\frac{\alpha}{R^{\frac{17}{4}}}-\frac{12}{25}\frac{1}{R^{\frac{11}{2}}}+\ldots,\\
&\nonumber B(r) \sim \frac{1}{4}\ln{R} + \frac{1}{4}\frac{1}{R^{\frac{5}{2}}}+\frac{\gamma}{R^{\frac{15}{4}}}+\frac{3}{8}\frac{1}{R^{5}}+\ldots,
\end{eqnarray}
where $R(r) = \Sigma r$.

We should remark here that the TN soliton exhibits exactly the same expansion with $\gamma = -\frac{1}{4}\sqrt{2}$, $\alpha = \frac{32}{25}\sqrt{2}$ and $\Sigma=\frac{1}{\sqrt{2l^2}}$. This feature enables us to define a mass of a WBH in terms of asymptotic integrals.
\subsection{Mass of the warped black hole}
The definition of a mass for the WBH is not straightforward because commonly known mass definitions (e.g. Komar mass \cite{komar}) rely on asymptotic flatness of a given solution. Yet, our spacetime is not asymptotically flat. It is locally asymptotically flat and we have to resort to other methods for mass computation.
Abbott and Deser \cite{masax} have proposed a general definition of conserved charges.  They replace asymptotically flat background with a spacetime of an arbitrary asymptotics which is supposed to be the vacuum of the system. Next, they use its isometries to generate conserved quantities.  Below we present an application of this method to the case in hand. 

We choose the TN soliton as a background solution and decompose the metric to factor-out the background:
\begin{equation}
g_{\alpha \beta}= \bar{g}_{\alpha \beta}+ h_{\alpha \beta}
\end{equation}
One can define
\begin{eqnarray}
H^{\alpha \beta} = h^{\alpha \beta} - \frac{1}{2}\bar{g}^{\alpha \beta} h, \\
K^{\alpha \beta \gamma \delta} = \frac{1}{2}\left(\bar{g}^{\alpha \delta} H^{\gamma \beta} + \bar{g}^{\gamma \beta} H^{\alpha \delta} - \bar{g}^{\alpha \gamma} H^{\beta \delta} - \bar{g}^{\beta \delta}H^{\alpha \gamma} \right).
\end{eqnarray}
All indices are raised and lowered using the background metric, $\bar{g}_{\alpha \beta}$. The TN soliton is static, therefore it possesses a Killing vector field $\bar{\xi}^{\alpha}\partial_\alpha = -\frac{\partial}{\partial t}$. The charge associated with this isometry can be interpreted as a mass and equals to
\begin{equation}
M_{AD} = \frac{1}{2 \pi^3} \oint \sqrt{-\bar{g}}\left[\bar{\xi}_\alpha \bar{\nabla}_\beta K^{t i \alpha \beta} - K^{t j \alpha i} \bar{\nabla}_j \bar{\xi}_\alpha \right] dS_i
\end{equation}
where $dS_i$ is the spatial surface area element at large radius, the index t denotes the time coordinate index. Greek indices run through all spacetime directions while Latin ones run through space directions. The integral should be performed at spatial infinity. That explains why this method is sometimes referred to as method of asymptotic integrals. We can, therefore, use the expansion \eref{wokolnieskonczonosci} for computation of the above integral to obtain
\begin{equation}
\label{masas}
M_{AD} = \frac{4}{5}\sqrt{2}- \frac{5}{4}\alpha - \frac{16}{5} \gamma.
\end{equation}
This result is valid only if the scaling parameter $\Sigma$ of the background TN soliton is the same as for WBH. Otherwise the integral is divergent. Note that $\Sigma$ does not appear in the above formula.

\begin{figure}[angle=90,htp]
\centering
\scalebox{0.3}{\includegraphics{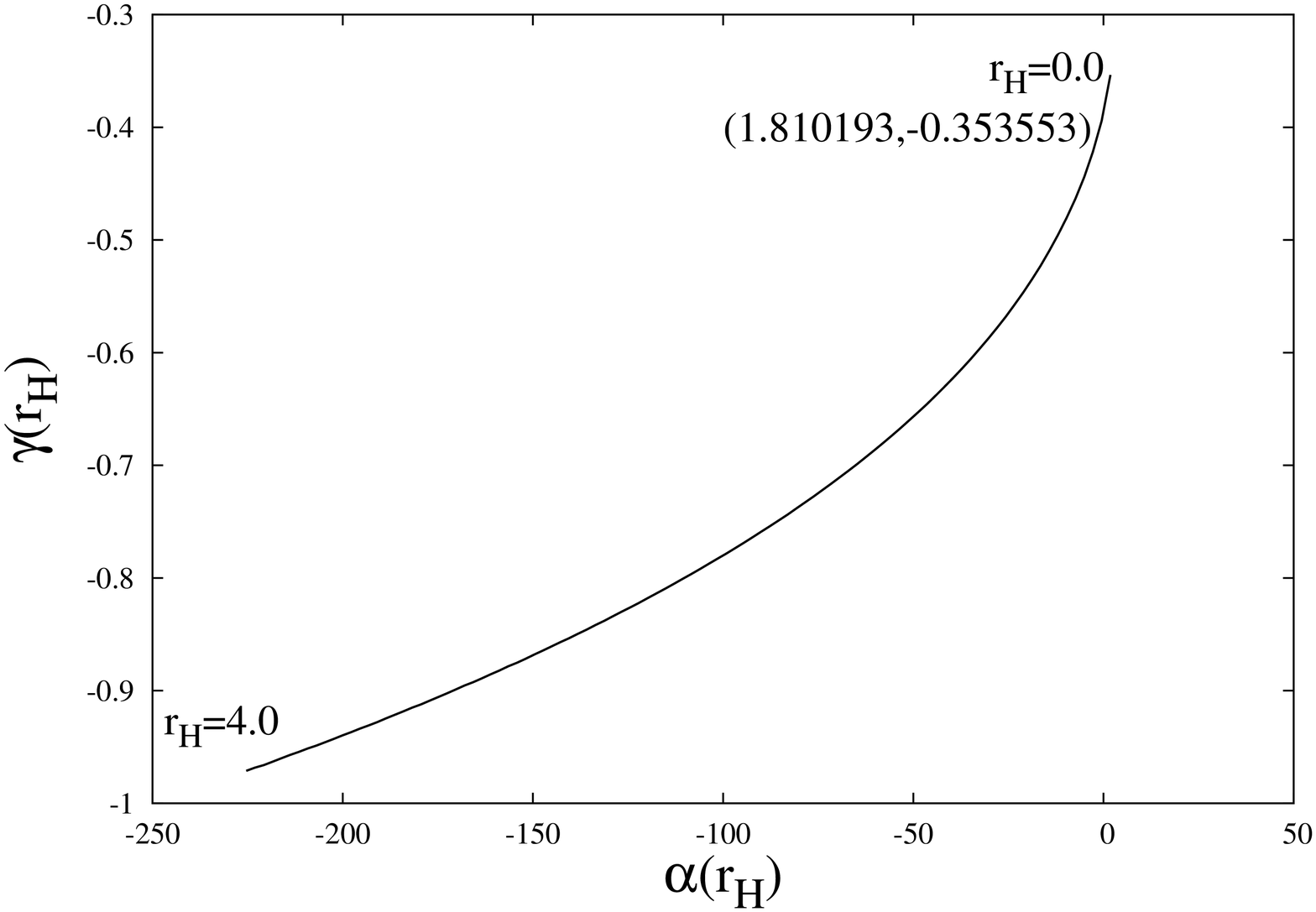}}
\caption{The relation between asymptotic parameters $\alpha$ and $\gamma$ for WBH is plotted as a parametric curve, horizon radius $r_H$ being the parameter. It can be seen that for $r_H \to 0$ the curve converges to $\{\gamma,\alpha\}$ of the soliton, $\gamma = -\frac{1}{4}\sqrt{2}$ and $\alpha = \frac{32}{25}\sqrt{2}$.}
\label{kobo}
\end{figure}

It is interesting that the WBH solution is parameterized by different number of parameters near $r_H$  than near infinity. More precisely, near $r_H$ we have two parameters $\{\beta,r_H\}$ while near infinity - three $\{\Sigma,\alpha,\gamma\}$. Therefore there must be a constraint involving $\Sigma, \alpha$ and $\gamma$ which reduces the freedom at infinity. We determined this constraint numerically as follows. We chose $r_H$, shot the solution to infinity and fine tuned $\beta$ to get $\Sigma=1$ from the fit. Then we were able to fit $\alpha$ and $\gamma$. The results are presented in \fref{kobo} as a curve parameterized by $r_H$. Combining this knowledge with the equation \eref{masas} we can compute the mass of the WBH basing solely on $r_H$. 

One might wonder what will happen if we perform shooting from infinity towards horizon and take $\alpha$ and $\gamma$ lying outside the curve from Figure \ref{kobo}. Using numerical simulations we looked at this case and observed that $A>0$ and  becomes singular near $r=0$.  This is not a surprise, since the reasoning presented in the proof of existence of WBH (e.g. $A$ can not cross $1$ from below) is only valid when shooting forward, not backward. 
 
\subsection{Comparison to the 4+1 dimensional case}
In the 4+1 dimensional model described in \cite{biz1} the black hole solutions are explicitly known. It is instructive to look at the parameter dependence between near horizon and far from horizon power series expansions and compare it to 6+1 dimensional case where only numerical relations are known. Let us begin by recalling the lower dimensional ansatz:
\begin{eqnarray}
ds^2=-Ae^{-2\delta}dt^2+ A^{-1}dr^2 + \frac{1}{4}r^2e^{2B}(d\theta^2+\sin^2{\theta}d\phi^2) \\
\nonumber \quad + \frac{1}{4} r^2 e^{-4B} (d\psi+\cos{\theta}d\phi)^2.
\end{eqnarray}
The $A$ and $B$ functions have an analogous interpretation as in 6+1 dimensional model.
From the above ansatz we get field equations. Below we present their static version
\numparts
\begin{eqnarray}
\label{autol1}
 r(rB')'  A  +\frac{2}{3} rB' 
 \left( 4\,{e^{-2\,B  }}-{e^{-8\,B  }}
 \right) -\frac{4}{3} \left( {e^{-2\,B  }} - {e^{-8\,B  
  }}\right)=0, \\
\label{autol2}
rA'  =-2\,A   \left( 1+
  r^2B'^{2} \right) +\frac{2}{3}\left(4
{e^{-2\,B }}-{e^{-8\,B  }}\right), \\
\label{autol3}
\delta' = -2rB'^2.
\end{eqnarray}
\endnumparts
where primes denote differentiation with respect to $r$. 
Next, we present KK solution
\begin{eqnarray}
B_0=\frac{1}{3} \ln(1+\frac{\rho}{m}), \quad A_0 = \frac{(1+\frac{4\rho}{3m})^2}{(1+\frac{\rho}{m})^{8/3}}, \quad e^{2\delta_0}=A_0,
\end{eqnarray}
where
\begin{eqnarray}
r=2m^{1/2}\rho^{1/2}(1+\frac{\rho}{m})^{1/6}
\end{eqnarray}
and $m>0$ is a free parameter which sets the scale (corresponding to $l$ of \eref{soliton}).
We also recall the black hole solution \cite{dziura4}:
\begin{eqnarray}
B=\frac{1}{3}\ln{\left(\frac{\Sigma}{2P}\right)},\quad A=\frac{\rho r^2 (3\Sigma + \Delta)^2}{36\Sigma^3\Delta^2}, \quad e^{2 \delta}=\frac{\Delta}{\rho}A
\end{eqnarray}
where 
\begin{eqnarray}
\Delta = \rho+3M-\sqrt{M^2+2P^2}, \quad \Sigma=\rho+M+\sqrt{M^2+2P^2}, \\
\nonumber   r=4^{2/3}P^{1/3}\Sigma^{1/6}\Delta^{1/2}.
\end{eqnarray}
Here $0<P\leq 2M$. To see that this is indeed a black hole we can find local asymptotics similar to \eref{rozw2}
\begin{eqnarray}
B(r) \sim \beta + \frac{2(e^{-2\beta}-e^{-8\beta})}{4e^{-2\beta}-e^{-8\beta}}\left(\frac{r}{r_H}-1\right),\\
\nonumber A(r) \sim \frac{2}{3}\left(4e^{-2\beta}-e^{-8\beta}\right)\left(\frac{r}{r_H}-1\right).
\end{eqnarray}
This time, though, we have explicit formulas connecting $\beta$ and $r_H$ with $P$ and $M$
\begin{eqnarray}
\beta = \frac{1}{3} \ln{\frac{M+\sqrt{M^2+2P^2}}{2P}},\\
\nonumber r_H^6=4^4P^2(M+\sqrt{M^2+2P^2})(3M-\sqrt{M^2+2P^2}).
\end{eqnarray}

Next,   careful asymptotic analysis applied to (\ref{autol1}-\ref{autol2}) gives us the following near-infinity expansions:
\begin{eqnarray}
\label{wokolnieskonczonosci-4}
&A(r) \sim \frac{16}{9}\frac{1}{R}+\frac{\alpha}{R^{\frac{5}{2}}}+\left({\frac {8}{9}}-\alpha\,\gamma+\frac{16}{3}\,{\gamma}^{2}\right)\frac{1}{R^{4}}+\ldots,\\
&\nonumber B(r) \sim \frac{1}{2}\ln{R} +  \frac{\gamma}{R^{\frac{3}{2}}} - \left( {\frac {9}{32}}\,\alpha\,\gamma+{\gamma}^{2}+\frac{1}{8}\right) \frac{1}{R^3} \ldots,
\end{eqnarray}
where $\alpha, \gamma$ are free parameters and $R=\Sigma r$ represents scaling freedom. In contrast with the 6+1 dimensional case, free parameters appear just after the leading order term. Note also, that the next order depends explicitly on these parameters. For KK monopole we get $\alpha = -\frac{16}{9}, \quad \gamma = \frac{1}{4}, \quad \Sigma = \frac{1}{2m}$. For the black hole we have
\begin{eqnarray}
\label{ooos}
\alpha = -\frac{8}{9} \frac{M+\sqrt{M^2+2P^2}}{4^{5/2}P^{7/2}}, \quad \gamma=-\frac{M - \sqrt{M^2+2P^2}}{{(4P)}^{5/2}}, \quad \Sigma=\frac{1}{4P} 
\end{eqnarray}
Therefore $P$ is simply the scale factor. It is evident that two parameters $\{M,P\}$ are translatable to $\{\beta, r_H\}$ on one hand and to $\{\alpha,\gamma , \Sigma \}$ on the other hand.  

To compare these results with the 6+1 dimensional case we set $P=\frac{1}{4}$ and get $\Sigma=1$.  One can now see that  $r_H$ depends solely on $M\geq\frac{1}{8}$ and that for $M=\frac{1}{8}$ we have $r_H=0$ - black hole degenerates to the KK monopole. $\beta$ is also fixed by the choice of $M$ so  we can parametrize the black hole up to scaling by $r_H$ alone. 

Next, elimination of $M$ from \eref{ooos} gives us a surprisingly simple relation between $\alpha$ and $\gamma$
\begin{equation}
\gamma = -\frac{4}{9} \frac{1}{\alpha}.
\end{equation}
It can be easily seen from Figure \ref{kobo} that in the 6+1 dimensional case the corresponding relation is very different.

In the end we present  mass of the 4+1 dimensional warped black hole computed using the same method as for 6+1 dimensional case:
\begin{eqnarray}
\label{masa4}
M_{4+1} = -\frac{1}{12}\left(8+32\gamma+9\alpha\right).
\end{eqnarray}
After substituting \eref{ooos}, we get
\begin{eqnarray}
M_{4+1}= \frac{16M- 2}{3}.
\end{eqnarray}

\section{Numerical results}
In order to investigate nonlinear perturbations we have resorted to numerical simulations.
Using second order finite difference code 
we have solved the equations (\ref{prozniowe}-\ref{ostatnie}) for several families of regular initial
data representing perturbations of Taub-NUT soliton (\ref{soliton}).
The overall picture does not depend on the specific choice of a family.
The results shown below were generated for initial data of the form (using momentum variable $P = e^\delta A^{-1} \dot{B}$):
\begin{eqnarray}
\label{poczatkowe}
B(0,r) = B_0(\rho(r)), & P(0,r)=p \left( \frac{r}{r_0}\right)^4 e^{\frac{(r-r_0)^4}{\sigma}}.
\end{eqnarray}
The scaling parameter of the initial soliton has been fixed to $l^2 = \frac{1}{2}$.
Parameters $\sigma$ i $r_0$ are fixed for given examples, whereas the amplitude $p$ varies.

\subsection{Small perturbations}
It has been observed that for small perturbations, by which we mean small value of the control parameter $p$, the perturbation is being radiated to infinity and the solution settles down to the Taub-NUT soliton with the same parameter $l$ as the perturbed one.
\begin{figure}[htb]
\begin{center}
\includegraphics[scale=1.0]{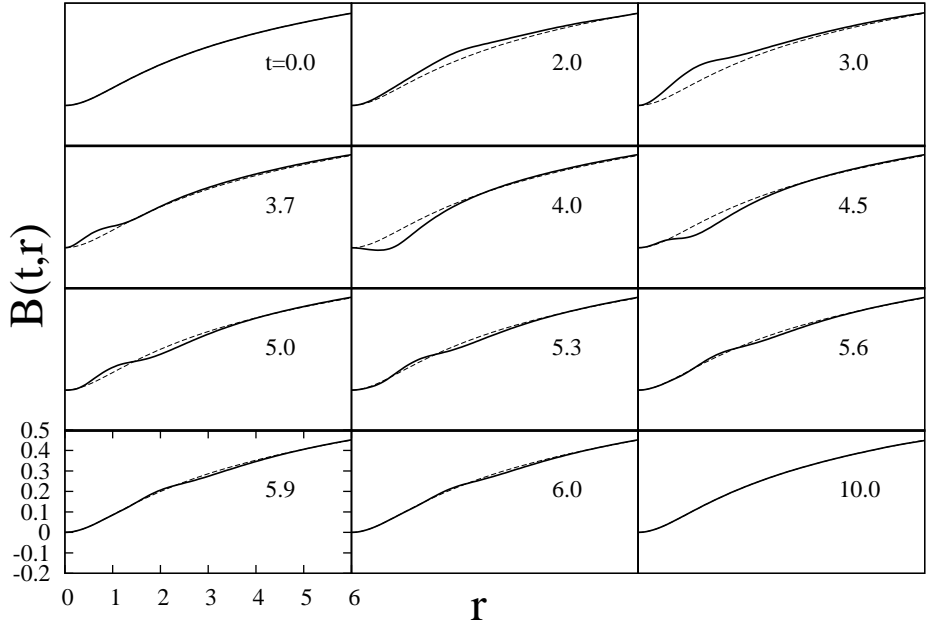}
\end{center}
\caption{Asymptotic stability of the Taub-NUT soliton in 6+1 spacetime dimensions. For initial data \eref{poczatkowe} with a small amplitude ($p=0.02$, $r_0=3$, $\sigma=1$) we plot a series of snapshots of the function B(t,r) (where t is central proper time). The dashed line shows the unperturbed TN soliton. During the evolution the excess energy of the perturbation is clearly seen to be radiated away to infinity and the solution returns to equilibrium.}
\label{solitonb}
\end{figure}

\begin{figure}[htb]
\centering
\includegraphics[scale=1.0]{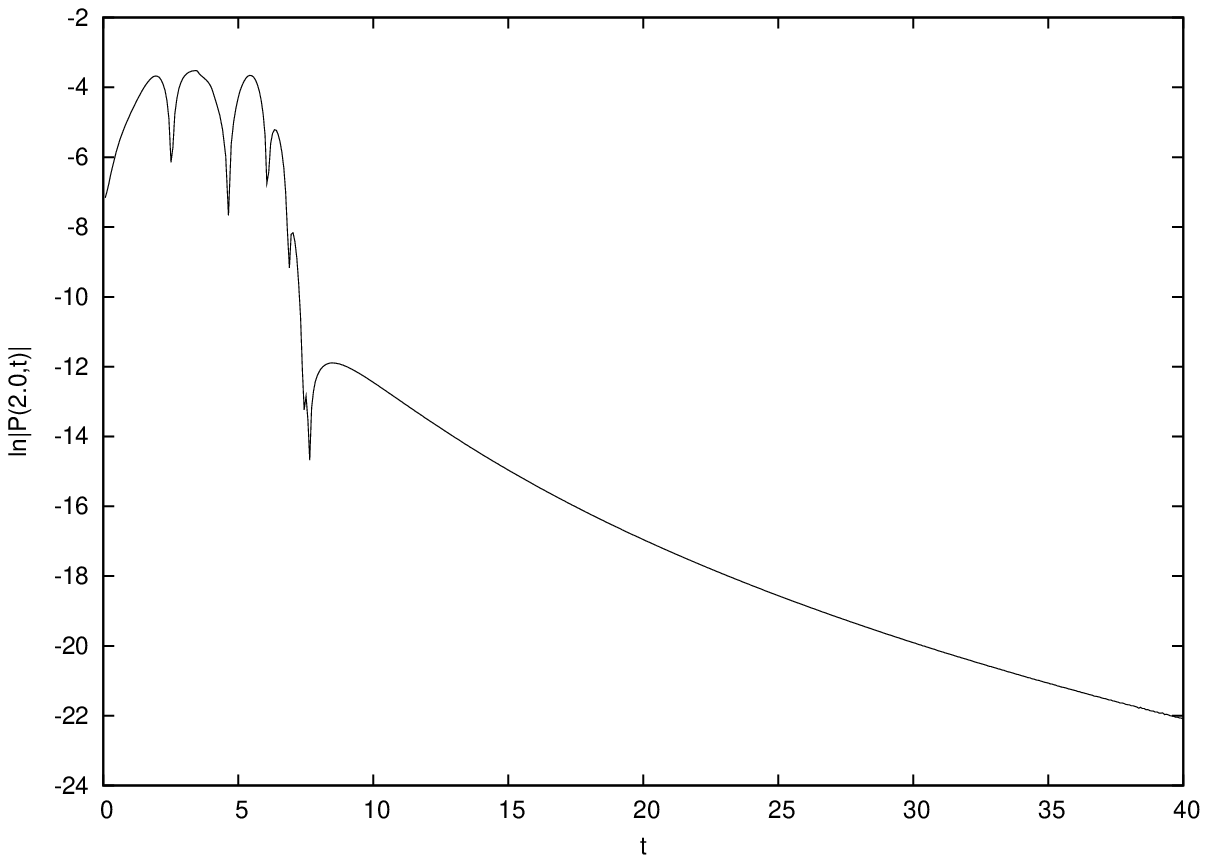}
\caption{The convergence to the TN soliton. For the same initial data as in Figure \ref{solitonb} we plot $\ln{|P(r_0,t)|}$ as a function of time for $r_0=2.0$. Asymptotic power-law tail can be seen.}
\label{pointwise}
\end{figure}

One might wonder why $l$ is not altered by the evolution. If we recall that $l$ is connected with the scaling constant by relation $\Sigma=\frac{1}{2l^2}$, we see that changes in $l$ would change scaling. The mass formula \eref{masas} becomes then divergent - it would require an infinite energy to change scaling, whereas our perturbation is of finite energy.  

\begin{figure}[htb]
\centering
\includegraphics[scale=1.0]{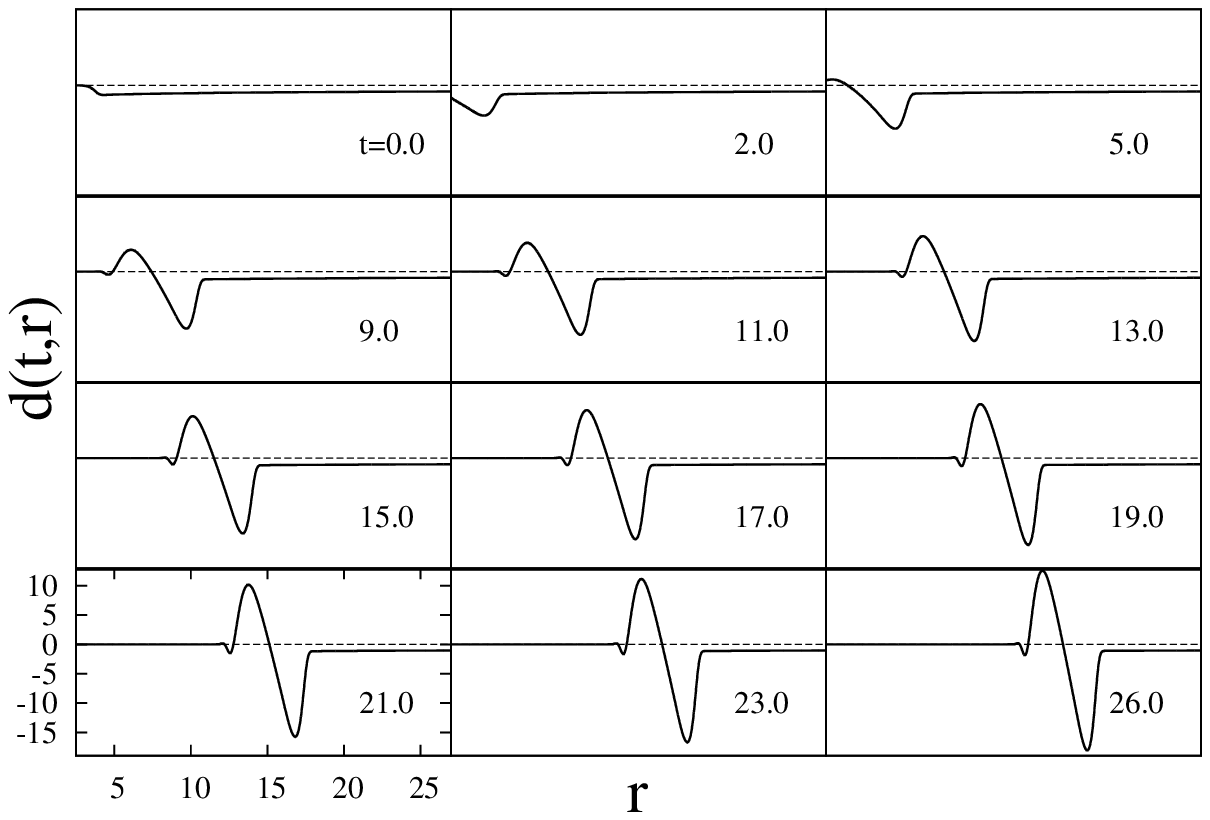}
\caption{Relaxation of the $r^{-4}$ term in asymptotic expansion. For the same initial data as in Figure \ref{solitonb} we plot series of snapshots for $d(t,r)$ as defined in \eref{rdoczwartejfun} (t is central proper time). The dashed line indicates zero. During the evolution the $r^{-4}$ coefficient goes to zero behind a wave travelling towards infinity.  }
\label{rdoczwartej}
\end{figure}

An interesting aspect of the relaxation process is connected with the absence of $r^{-4}$ term in expansion (\ref{wokolnieskonczonosci}) of $A$.
If one examines \eref{prozniowe} closely it can be seen that the compact-support perturbation imposed on the momentum variable instantly introduces $r^{-4}$ term in the expansion of $A$ for large r. If the perturbed soliton is really stable, this term should eventually vanish. More precisely, from \eref{wokolnieskonczonosci} we know that the function
\begin{equation}
\label{rdoczwartejfun}
d(t,r) = r^4 \left(A(t,r) -  A_0(r) \right)
\end{equation}
represents the coefficient in front of $r^{-4}$ term. 
Figure \ref{rdoczwartej} presents the behaviour of this function. 

\subsection{Late time tail}

In order to better understand the relaxation process we transformed the linear equation \eref{strum} into single mode wave equation
\begin{equation}
-\frac{d^2}{dx^2}u(x) + V(x) u(x) = \lambda^2 u(x) 
\end{equation}
using transformation
\begin{eqnarray}
v(\rho) = {\frac {\sqrt [4]{{\rho}^{3} \left( \rho+4 \right) ^{3}}}{{\rho}^{2}\sqrt {\rho+2} \left( \rho+4 \right) }} u, \\
d x = \sqrt{\frac{(\rho+2)^2}{2\rho(\rho+4)}}d \rho.
\end{eqnarray}
The potential has the following asymptotic behaviour
\begin{eqnarray}
V(x) = \frac{2}{x^2}+\frac{2\sqrt{2}}{x^3}+\frac{6}{x^4} + O(\frac{1}{x^5}).
\end{eqnarray}
According to work of Ching et al. \cite{ching} and Bizoń et al. \cite{biz5} the solution should exhibit asymptotic power-law tail
\begin{equation}
\label{ogg}
u(x,t) \sim \frac{1}{t^6}.
\end{equation}
We are aware of the possible existence of nonlinear tail, however our numerical simulations are in good agreement with \eref{ogg}. Furthermore, the amplitude of the tail scales linearly with initial data, which suggests that even if nonlinear tail is present it decays faster than the linear one.

Our simulations strongly suggest that the TN soliton is nonlinearly stable against small perturbations.  
However, just like its lower dimensional cousin, it is being destabilized by sufficiently strong perturbations. Warped black holes are important in this process.

\subsection{Large perturbations}
We have found that for sufficiently strong perturbations the inner part of the soliton collapses and horizon is being formed for $r = r_H>0$. The external part of the solution settles down to a WBH. The scaling parameter $l$ is preserved.  
\begin{figure}[h]
\centering
\includegraphics[scale=1.0]{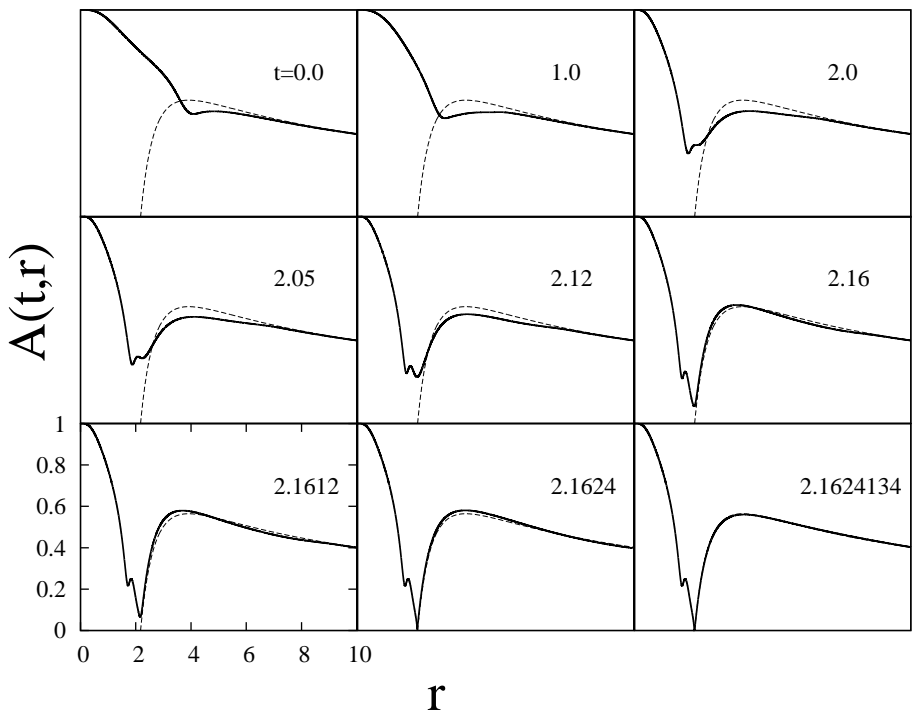}
\caption{Instability of the Taub-NUT soliton for large perturbations. For initial data \eref{poczatkowe} with large amplitude ($p=0.1$,$r_0=3$, $\sigma=1$) we plot a series of snapshots of the function A(t,r) (where t is central proper time). During the evolution A(t,r) drops to zero near $r=r_H \simeq 2.174$ which signals the formation of a horizon there. Outside the horizon the solutions settles down to the warped black hole (dashed line) with  $M=0.612$. }
\label{dziurka}
\end{figure}

\section{Conclusions}
W have shown that the TN soliton in 6+1 dimensions is classically stable just like the KK monopole. Therefore the stability is neither dimension specific nor supersymmetry dependent feature and its origin should be sought elsewhere. We have also shown that for sufficiently strong perturbations a collapse to the warped black hole occurs. The black hole properties have been studied in detail and the proof of global existence has been presented. Still, there are open questions concerning the model. The first one is connected with the critical behaviour on the threshold of black hole formation. In 4+1 case the discrete type II self similarity was observed and it might be interesting to compare those results with the observations concerning the case in hand. The other interesting feature that can be examined is the appearance of quasinormal modes in asymptotic pointwise convergence presented in the Figure \ref{pointwise}.
\ack
I want to express my gratitude to Piotr Bizoń for hours of enlightening discussions. This work was supported in part by the Polish Ministry of Science grant no. N N202 079235 (2008-2010).

\appendix
\section{}
This appendix is devoted to the formal proof of the existence theorems 
for static equations. Let us begin with the theorem which has been proven in \cite{proof} (Proposition 12):
\begin{thm}
\label{istnienie}
Suppose a system of differential equations for m+n functions
$u = (u_1 , . . . , u_m )$ and $v = (v_1 , . . . , v_n )$,
\begin{eqnarray}
\label{postac}
t \frac{du_i}{dt} = t^{\mu_i} f_i(t,u,v), & t \frac{dv_i}{dt} = -\lambda_i v_i + t^{\nu_i} g_i(t,u,v),
\end{eqnarray}
with constants $\lambda_i$ with $Re(\lambda_i) > 0$ and integers $\mu_i $,
$\nu_i \geq 1$ and let $\mathfrak{C}$ be an open subset of $R^m$ such that the functions 
f and g are analytic in a neighbourhood of $t = 0$, $u = c$, $v = 0$ for all 
$c \in \mathfrak{C}$. Then there exists a unique m-parameter family of solutions of the system (\ref{postac}) 
with boundary conditions $(u, v) = (c, 0)$ at $t = 0$, defined for $c \in \mathfrak{C}$ and 
$|t| < t_0 (c)$ with some $t_0 (c) > 0$. These solutions satisfy
\begin{eqnarray}
\nonumber u_i(t)=c_i+ O(t^{\mu_i}), & v_i(t)=O(t^{\nu_i}),
\end{eqnarray}
and are analytic in t and c.
\end{thm}
Now, we cast static equations into the form:
\begin{eqnarray}
\label{stat1}& \left(A r^5 B' \right)' = -4r^6AB'^3 + \frac{6}{5}r^3\left(e^{-2B}-e^{-12B}\right), \\
\label{stat2}& A' = \frac{-4A}{r}\left(1+r^2B'^2\right) + \frac{1}{5r}\left( 24e^{-2B}-4 e^{-12B} \right),\\
\label{stat3}& \delta' = -4rB'^2.
\end{eqnarray}
The first two equations are independent of the third. This allows us 
to consider only (\ref{stat1},\ref{stat2}).
\subsection{Existence near $r=0$}
We will try to prove the relations (\ref{wokolzera}).
In order to do this, we perform a transformation:
\begin{eqnarray}
\nonumber \omega_1 = \frac{1-e^{-2B}}{2r^2},& \omega_2=\frac{AB'}{r},&\omega_3=\frac{1-A}{r^4}.
\end{eqnarray}
According to our supposition on local behaviour we should get:
\begin{eqnarray}
\nonumber \omega_1(r) \sim b, & \omega_2(r) \sim  2b, & \omega_3(r) \sim 8b^2.
\end{eqnarray}
Now, equations (\ref{stat1},\ref{stat2}) take the form:
\begin{eqnarray}
&\nonumber r\omega_1' = -2\omega_1 + \omega_2 + r^2f_1, \\ 
&r\omega_2' = 12\omega_1-6\omega_2 + r^2f_2, \\
&\nonumber r\omega_3' = 4\left( \omega_2^2 + 12  \omega_1^2 - 2 \omega_3\right) + r^2f_3,
\end{eqnarray}
where $f_i$ are polynomials in $\omega, \frac{1}{A}, r^2$. We need to perform another transformation:
\begin{eqnarray}
&\nonumber \omega_1 = u_1 - v_1, \\ 
&\omega_2 =  2u_1 + 6v_1,\\
&\nonumber \omega_3 = v_2+ 8(3v_1^2 + u_1^2),
\end{eqnarray}
to finally get 
\begin{eqnarray}
&\nonumber ru_1' = r^2g_1, \\ 
&r v_1' = -8 v_1+ r^2h_1, \\
&\nonumber r v_2' = -8 v_2 + r^2 h_2,
\end{eqnarray}
where g, h are polynomials in $u,v,\frac{1}{A},r^2$. This system has the form (\ref{postac}), thus from Theorem \ref{istnienie} we acquire
\begin{thm}
\label{wokolzera0}
There exists an unique, one-parameter family of regular solutions for (\ref{stat1}-\ref{stat2}) near $r=0$ analytic in $b$ and $r$ such that
\begin{eqnarray}
\nonumber & B(r) = br^2+O(r^4),\,\,\,\,  A(r) = 1-8b^2 r^4 + O(r^8), \\
\end{eqnarray}
\end{thm}
\subsection{Existence near $r_H: A(r_H)=0$}

We need a transformation which casts (\ref{stat1}-\ref{stat2}) to the form (\ref{postac}), around $r_H>0 : A(r_H)=0$:
\begin{eqnarray}
\rho = r- r_H, & u_1(\rho) = r, & u_2(\rho) = \frac{1-e^{-2B}}{2},\\
\nonumber \omega_1 = \frac{A}{\rho},& \omega_2=\frac{AB'}{\rho}.
\end{eqnarray}
We get
\begin{eqnarray}
&\nonumber \rho u_1' = \rho, \\ 
& \rho u_2' =\rho \frac{\omega_2}{\omega_1}, \\
&\nonumber \rho \omega_1' = -\omega_1 + F_1(\rho) + \rho f_1,\\
&\nonumber \rho \omega_2' = -\omega_2 + F_2(\rho) + \rho f_2,
\end{eqnarray}
where $f_1$ and $f_2$ are polynomials in $\omega_1, \omega_2, \frac{1}{\omega_1}, \frac{1}{u_1}$  and
\begin{eqnarray}
F_1(\rho) = \frac{1}{5u_1} \left( 24(1-2u_2) - 4 (1-2u_2)^6 \right),\\
\nonumber F_2(\rho) = \frac{6}{5u_1^2} \left(1-2u_2 -  (1-2u_2)^6 \right).
\end{eqnarray} 
Next, transformations
\begin{eqnarray}
\omega_i = v_i + F_i
\end{eqnarray}
give :
\begin{eqnarray}
\rho v_i = -v_i + \rho g_i(u,v),
\end{eqnarray}
where $g_i$ are polynomials in $\omega_1, \omega_2, \frac{1}{\omega_1}, \frac{1}{u_1}$. Those polynomials are regular around $u_1 = r_H,  u_2 = \beta, v_i = 0$.
According to Theorem \ref{istnienie} we get:
\begin{thm}
\label{istnienie3}
There exists an unique two-parameter family of regular solutions of the system (\ref{stat1}-\ref{stat2}) around $r_H : A(r_H)=0$, $r_H>0$,  analytic in $\beta$, $r_H$ and $r$, so that:
\begin{eqnarray}
\nonumber B(r-r_H) = \beta + O(r-r_H), & A(r-r_H) = O(r-r_H).
\end{eqnarray}
\end{thm}
\subsection{Existence near infinity}
Let us consider a coordinate change $ u = \frac{1}{r^{1/4}} $.
We then take $ \rho = \Sigma u$, $\Sigma > 0$: 
\begin{eqnarray}
\nonumber && A(\rho) = \frac{32}{25} (\rho^2-\rho^{12}) + \mathfrak{A}(\rho) \rho^{17},   \\
\nonumber && B(\rho) = -\ln{\rho} + \frac{1}{4} \rho^{10}  + \mathfrak{B}(\rho) \rho^{15},   \\
\nonumber && S(\rho) = \frac{\rho}{u}. \\ \nonumber
\end{eqnarray}
Some more transformations are required:
\begin{eqnarray}
\nonumber && e^{-\frac{1}{2}\rho^{10}(1+4 \rho^5 \mathfrak{B}(\rho))} = 1 -\frac{1}{2}\rho^{10}(1+4 \rho^5 \omega(\rho)),\\
\nonumber && Z(\rho) = \frac{d \omega}{d \rho}. \nonumber
\end{eqnarray}
Those transformations leave us with the set of equations:
\begin{eqnarray}
\nonumber &&\rho \frac{d \mathfrak{A}}{d \rho} = \rho f(\mathfrak{A},\omega,S,Z,\rho), \\
\nonumber &&\rho \frac{d S }{d \rho} = \rho ~ 0 ,  \\
\nonumber &&\rho \frac{d \omega }{d \rho} = \rho Z(\rho), \\
\nonumber &&\rho \frac{d Z }{d \rho} = -16 Z(\rho) + \rho g(\mathfrak{A},\omega,S,Z,\rho), \nonumber
\end{eqnarray} 
where $f,g$ are regular for $ \omega \to 0, \mathfrak{A} \to \alpha, \omega \to \gamma, S \to \Sigma > 0, \rho \to 0$.
Those equations are of the form (\ref{postac}). We now get from Theorem \ref{istnienie}:

\begin{thm}
\label{istnienie4}
There exists an unique three-parameter family of regular solutions for the system (\ref{stat1}-\ref{stat2}) around infinity with asympotics $B(r) \sim \ln{r}$, analytical in $\Sigma > 0$, $\alpha$ and $\gamma$, so that:
\begin{eqnarray}
\label{wokolnieskonczonosci2}
&A(r) = \frac{32}{25}\frac{1}{\sqrt{R}}-\frac{32}{25}\frac{1}{R^3}+\frac{\alpha}{R^{\frac{17}{4}}}+ O(\frac{1}{R^{18/4}}),\\
&\nonumber B(r) = \frac{1}{4}\ln{R} + \frac{1}{4}\frac{1}{R^{\frac{5}{2}}}+\frac{\gamma}{R^{\frac{15}{4}}}+O(\frac{1}{R^4}),
\end{eqnarray}
where $R = \Sigma r$.
\end{thm}

\section{}
We present an explicit form of $V(\rho)$ introduced in the proof
of linear stability :
\begin{eqnarray}
\nonumber \fl V(\rho)= \frac{4}{5}\frac{1}{\rho^5 l^2 (\rho+2l)^4(\rho^2+5\rho l+5l^2)^2(\rho+4l)} \times \Big(3 l^2 \rho^{12} + 60 l^3 \rho^{11} + 502 l^4 \rho^{10} + \\
\nonumber \fl+\quad 2250 l^5 \rho^9 + 5665 l^6 \rho^8 + 7500 l^7 \rho^7 + 36 \rho^6 P^6 Q^4 - 3 \rho^6 P Q^4 + 4000 \rho^6 l^8 - 42 \rho^5 P Q^4 l + \\
\nonumber \fl+ \quad 504 \rho^5 P^6 Q^4 l - 237 \rho^4 P Q^4 l^2 + 2844 \rho^4 P^6 Q^4 l^2 - 690 \rho^3 P Q^4 l^3 + 8280 \rho^3 P^6 Q^4 l^3 + \\
\nonumber \fl+ \quad 13140 \rho^4 P^6 Q^4 l^2 - 1095 \rho^2 P Q^4 l^4 - 900 \rho P Q^4 l^5 + 10800 \rho P^6 Q^4 l^5 - 300 P Q^4 l^6 + 3600 P^6 Q^4 l^6 \Big)
\end{eqnarray}
where
\begin{eqnarray}
\nonumber P = \left( \frac{l^2(\rho+4l)}{(\rho+2l)^3}\right)^{\frac{1}{5}},\\
\nonumber Q = \left( l^2 \rho^5 (\rho+2l)^2 (\rho+4l) \right)^{\frac{1}{5}}.
\end{eqnarray}
\section*{References}

\bibliography{nut}

\end{document}